\documentstyle []{article}

\newcommand{\vi}{\hat{\mbox{\boldmath $\i$}}}
\newcommand{\vj}{\hat{\mbox{\boldmath $\j$}}}
\newcommand{\vk}{\hat{\mbox{\boldmath $\rm k$}}}

\newcommand{\nab}{\mbox{\boldmath $\nabla$}}

\begin{document}
 
\title{The hydrodynamical response of a tilted circumbinary disc:
linear theory and non--linear numerical simulations}
\author{John D. Larwood and John C.B. Papaloizou}

\maketitle

\centerline{\bf Abstract}

\noindent In this paper we present an analytical and numerical study of the
response of a circumbinary disc subject to the tidal--forcing of a binary
with a fixed circular orbit. We consider isentropic fluid discs with a range
of thicknesses and binaries with a range of mass ratios, orbital separations
and inclination angles.  

\noindent Our numerical simulations are implemented using an SPH code such that
we can consider the hydrodynamics fully in three--dimensions. For our
unperturbed disc models, we find the numerical shear viscosity to be equivalent
to a constant kinematic viscosity and we calibrate its
magnitude. Writing a scaling relation for the shear viscosity manifest in our
models, we deduce that the disc thickness cannot be varied without affecting
the viscosity in these kinds of SPH disc models, as is supported by our
numerical results.

\noindent It is found that maintainance of an inner cavity owing to the tidal
truncation of the disc is effective for non--zero orbital inclinations. Also
we show that our model discs may precess approximately like rigid bodies,
provided that the disc is able to communicate on a length scale comparable to
the inner boundary radius by either sonic or viscous effects, 
in a sufficiently small fraction of the local precession period. It is found
also that the surface density in the disc should not decrease too rapidly
with increasing radius, otherwise the disc may separate into disconnected
annuli. Furthermore, the disc precession period may tend to infinity if the
disc outer edge is allowed to become arbitrarily large, the disc suffering
only a modest quasi--steady warp near the inner boundary.

\noindent When the disc response is linear, or weakly non--linear, the
precession periods and the forms of warping that we measure in our numerical
results yield reasonable quantitative agreement with the analytical expressions
that we derive from a linear response calculation. For a stronger disc response
the results can agree poorly with our linear analysis, although some
qualitative features of the response remain intact. We demonstrate that the
response of a disc of non--interacting particles is qualitatively different,
showing a much larger kinematic disturbance and an ultimate global thickening
of the disc. 
 
\noindent This work is of relevance to a number of astrophysical phenomena
of current interest in star and planet formation; these include tidal
truncation and gap formation in accretion discs and the observational
characteristics of some young stellar objects.

\section{Introduction}

Direct imaging of protostellar discs in the Orion nebula (O'Dell et al. 1993,
McCaughrean \& O'Dell 1996) confirms the long--time hypothesis of their
role in low mass star formation. Other studies have shown
that binarity in local star--forming regions has a frequency in excess of that
normally attributted to solar type main sequence stars in the field.
This is found to apply over a wide range of orbital separations (for a review,
see Mathieu 1994). Furthermore, it is clear that
binary separations in these systems can be of the same order as, or {\it less}
than typical disc sizes. This is suggestive of the existence of
circumbinary discs as well as circumstellar discs and raises the question of
disc evolution under the influence of bodies other than a central mass alone.

\noindent Recent observations have clearly established that protostellar
discs can exist in binary and multiple systems. For
example, millimetric measurements made by Dutrey et al. (1994) and direct
imaging with {\it adaptive optics} by Roddier et al. (1996), reveal the
young binary GG Tauri as occupying a cavity defined by a circumbinary ring
of emission, thought to be scattered light from the inner edge of a much
larger circumbinary disc. Mathieu et al. (1996, in preparation) have made
observations of the remarkable multiple system UZ Tauri. They find that
this {\it quadruple} system consists of a sub-au binary with a circumbinary
disc, seperated by 500 AU from a 50 AU binary, with at least one of its
components carrying a circumstellar disc. Jensen et al. (1996) in a study
of $85$ systems in Scorpius--Ophicus and Taurus--Auriga find that the
sub--millimeter fluxes from binaries with seperations less than about
$100$ AU is generally less than that for wider or single systems, which they
suggest is due to star--disc interactions. Those authors place an upper limit
on circumbinary disc masses of $0.005 M_\odot$. 

\noindent Evidence for non coplanarity of binary orbits and disc planes is
provided by the low inclinations of the gas giants in the solar system.
Furthermore it is possible to model a small
asymmetry in the inner regions of the dusty disc around Beta Pictoris (detected
with the {\it Hubble Space Telescope}, unpublished) as a warp generated by a
low mass body on an inclined orbit interior to the disc (Mouillet et al.
1996). Other evidence comes from the third component of the young triple system
TY CrA apparantly in an orbit at a high inclination to
that of the inner binary component (Corporon et al. 1996), and the apparent
precession seen in protostellar jets (Bally \& Devine 1994). Also Terquem \&
Bertout (1993, 1996) considered the
observational appearence of a thin protostellar disc with warped geometry and
were able to synthesize a range of observed spectral energy distributions
(SEDs).

\noindent The above is suggestive of
non coplanar effects which should be seen in other such binary, and 
multiple systems. But relative inclinations prove difficult to extract from the
observational data available at present and the evidence for non coplanarity to
date has been mostly circumstantial in character. Some of the effects we
discuss in this paper may become accessible to observation in the near
future.

\noindent The tidal influence of a companion in a circular orbit upon an
accretion disc has
been extensively studied analytically in the linear regime and also in the
non--linear (as well as the linear) regime by numerical simulations (see Lin \&
Papaloizou 1993, and references therein). Both circumstellar and circumbinary
discs have been considered.  Artymowicz \& Lubow (1994) modelled the coplanar
response of a circumbinary disc to the tide of a binary in an orbit
with non--zero eccentricity. However, efforts have focussed for the most part
on coplanar configurations. 

\noindent The tidal disruption of a disc, by encounters with unbound
companions at arbitrary inclinations to the disc plane, has been shown to
result in a change in the relative inclination of the target disc (Heller
1993, Clarke \& Pringle 1993). 

\noindent Papaloizou \& Terquem (1995) calculated the linear response of a
fluid disc to the tide of a binary companion on an inclined circular orbit
exterior to a circumprimary disc. They showed that for a weak enough tidal
perturbation, the disc would be expected to  precess like a rigid body.
Larwood et al. (1996, hereafter LNPT) studied the same problem numerically with
a fully non--linear treatment in three spatial dimensions using SPH.
They were able to demonstrate disc warping and precession in the tidal
field of the companion. The conclusion was that the disc remained only
modestly warped, precessing approximately like a rigid body, provided that
the sound--crossing time in the disc was a sufficiently small fraction of the
precessional period. 

\noindent In this paper we extend the work presented in LNPT to consider a
circumbinary disc with an interior orbiting companion on a fixed circular
trajectory at an arbitrary inclination to the disc mid--plane. As in LNPT
we study the tidal interaction, disc truncation, warping, and
precession as a function of disc thickness, binary mass ratio and
orbital inclination.

\noindent In Section 2 we review the basic concepts of the theoretical
model we employ for our circumbinary disc. In Section 3 we
describe and discuss the numerical method we use to perform the simulations.
In Section 4 we discuss the linear tidal response of a disc to a binary
companion in an inclined orbit. The linear theory we use to describe the form
of tidally--induced warps in the thin disc limit is described in
Papaloizou \& Lin (1995a). Here we perform a linear calculation of the response 
to the secular terms in the perturbing potential that lead to
the warping and precession of the disc. In Section 5 we
describe the setting--up of the numerical simulations and some of the
diagnostics used. In Section 6 we give an account of our results. Finally, in
Section 7 we discuss our findings and give our conclusions.

\section{Basic Equations}

We consider the hydrodynamic equations governing the time 
evolution of a viscous gaseous medium. Assuming the disc self-gravity to be
negligible in comparison to the tidal field of the binary components, 
the governing equations are the equations of continuity and motion,
which may be written

\begin{equation}  
{{\partial \rho} \over {\partial t}} + \nab \cdot ( \rho {\bf v} )
 = 0,
\label{cont} 
\end{equation}

\begin{equation}
{{\partial {\bf v}} \over {\partial t}} + ( {\bf v} \cdot \nab ){\bf v} =
- \frac{1}{\rho} \nab P - \nab \Psi + {\bf f}_v, 
\label{dvdt} 
\end{equation}

\noindent where $\rho$, ${\bf v}$ and $P$ represent the fluid density,
velocity and pressure fields respectively. The gravitational
potential is $\Psi$, and ${\bf f}_v$ is the viscous force per
unit mass.

\noindent Here we shall ignore the details of heat generation and
transport and adopt a simple polytropic equation of state:

\begin{equation}
P= K \rho^{1+1/n},
\label{stat}
\end{equation}

\noindent where $n$ is the polytropic index and $K$ is
the polytropic constant. For all models considered here we take $n=1.5.$
The associated barotropic sound speed, $c_s,$ is given by

\begin{displaymath}
c_s^2 = \frac{(n+1)K}{n} \rho^{1/n}.
\end{displaymath}

\noindent We note that a constant $K$ specifies isentropy for the system
and thus any dissipated energy must be assumed to be radiated away.

\subsection{Equilibrium Disc Model}

In standard notation, the lowest order equation of hydrostatic equilibrium in
the vertical direction, for the unperturbed disc is 

\begin{equation} {1 \over \rho}{\partial P \over \partial z}= -\Omega^2 z ,
\label{vert}
\end{equation}
where $\Omega$ is the angular velocity.

\noindent For a disc satisfying the polytropic equation of state (\ref{stat}),
integration of (\ref{vert}) gives

\begin{equation} \rho =
\left({\Omega^2H^2\over 2K(n+1)}\right)^n(1-z^2/H^2)^n,
\label{vertrho}
\end{equation} 

\noindent where $H$ is the total vertical semi--thickness which
may be a function of $r$, and $\Omega$ is the disc rotation velocity. Hence
the surface density $\Sigma$:

\begin{equation}
\Sigma=c_n H\left({\Omega^2H^2\over 2K(n+1)}\right)^n,
\label{vertsig} 
\end{equation}

\noindent with $c_n = \Gamma(1/2)\Gamma(n+1)/\Gamma(n+3/2),$ where $\Gamma$
denotes the Gamma function. We also note from equation~(\ref{vertrho}) that

\begin{equation}
{\cal M} = \sqrt{2n}{\left( \frac{H}{r} \right)}^{-1}
\end{equation}

\noindent where ${\cal M} \equiv r \Omega / c_{s}$ is the Mach number,
evaluated at the disc mid--plane.

\noindent Finally we invoke the standard $\alpha$--prescription due to
Shakura \& Sunyaev (1973) to parameterise an anomalous viscosity. 
 Thus $\nu = \alpha c_{s} H$ gives the kinematic viscosity
coefficient $\nu$.

\section{Numerical Method}

\noindent The numerical method we employ to study the dynamics
of circumbinary accretion discs is a version of SPH
(see Monaghan 1992, and references therein)
developed by Nelson \& Papaloizou (1993, 1994). The reader is referred to these
papers for a description and standard tests of the method.
The formulation uses a variable {\it smoothing length} associated with
each particle, which is a function of the particle coordinates, defined in
such a way as to ensure accurate energy consevation. Using this method, a
particle's smoothing length is defined to be half of the mean distance of the
six most distant nearest neighbouring particles chosen from a list
of forty--five members at each time--step.

\noindent In order to stabilize the calculations in the presence of shocks
an artificial viscous pressure term is included, according to the prescription
of Monaghan \& Gingold (1983) as in LNPT. Although designed to prevent
particle penetration while giving positive definite dissipation, the practical
implimentation of the artificial viscosity
introduces a shear viscosity which results in disc spreading
much as in the standard theory of accretion
discs (Lynden-Bell \& Pringle 1974). This effect is quantifiable and
utilised in our calculations.

\subsection{Viscosity Calibration}

\noindent In LNPT we described how the shear viscosity
present in our model discs was calibrated by
comparing SPH calculations of the evolution of an unperturbed accretion disc
to solutions obtained by solving the well known diffusion equation for the
evolution of the surface density in a thin axisymmetric disc
(see Pringle 1981, for a review)
 
\begin{equation}
{\partial \Sigma \over \partial t} =
-{1 \over r}{\partial \over \partial r} \left\{
\left[ \left( r^2 \Omega \right)' \right]^{-1} 
{\partial \over \partial r}
\left( \nu \Sigma r^3 \Omega' \right) \right\}, 
\label{diffe}
\end{equation}

\noindent The primes in (\ref{diffe}) denote differentiation with respect to
$r$.

\noindent For the standard Monaghan \& Gingold (1983) parameters we
take (LNPT, see also Artymowicz \& Lubow 1994) the numerical
viscosity (defined locally) is essentially $\propto  0.5 c_{s} h$, where
$h$ is the local smoothing length. Since our smoothing length is spatially
adaptive $h \propto \rho^{-1/3}$
which cancels the density dependence of the sound speed for polytropic
models with $n=1.5$. Solving (\ref{diffe}) for $\nu=constant$ confirms
our naive expectation for constant kinematic viscosity to be manifest
in these model discs. For constant Mach number discs, the scaling for the
effective Shakura--Sunyaev $\alpha$ is then given by

\begin{equation}
\alpha \equiv {\nu {\cal M}^2\over r^2 \Omega}
\sim 0.025 (R_o/r)^{1/2}({\cal M}/10)^{2/3},
\label{visc}
\end{equation}

\noindent where $R_o$ is the radius of the outer boundary of the disc
and we have ignored a weak dependence on the total number of particles. 
So typically we find for ${\cal M}\sim 10$ a Reynolds' number
$(\equiv r^{2} \Omega / \nu) \sim 4000$, near the outer edge of the disc.

\noindent We note that this yields the effective Shakura--Sunyaev kinematic
viscosity $\propto {\cal M}^{-4/3}$. It is important to note that in these 
SPH disc models the disc thickness and viscosity may not be varied
independently. Thick discs are inevitably very viscous (see also
Artymowicz \& Lubow 1994, 1996). Simulation of thick discs with small viscosity
has to be performed with other numerical methods.

\subsection{The Gravitational potential}

\noindent We assume that the disc circulates around
a binary system, the orbit of which
is circular with angular frequency ${\bf \omega}.$
In addition we suppose that the orbital plane has an initial 
inclination angle $\delta$ with respect to the Cartesian $(x,y)$ plane, which
corresponds to the initial mid--plane of the disc. The $z$ axis is
normal to the initial disc mid--plane. We denote the unit
vectors in each of the Cartesian coordinate directions by $\vi$, $\vj$ and
$\vk$ respectively and the associated cylindrical polar coordinates are
$(r,\varphi,z).$ 

\noindent For a disc with negligible mass the plane of the orbit is
invariable and does not precess. When the mass of the secondary is very much
less than the mass of the primary, it is possible to have the origin of the
non rotating coordinate system  at the the primary. The
coordinates  and origin of time are such that the line of
nodes coincides with, and the secondary is on, the $x$ axis at $t=0,$ the
position vector ${\bf D}$ of the secondary is given as a function of time by

\begin{equation}
{\bf D} = D \cos \omega t \;  \vi +
D \sin \omega t \cos \delta  \; \vj +
D \sin \omega t \sin \delta \; \vk .
\label{D}
\end{equation}

\noindent where $D=|{\bf D}|$.

\noindent The total gravitational potential $\Psi$ at a point with position
vector ${\bf r}$ is given by

\begin{equation}
\Psi = - \frac{GM_p}{\mid \bf{r} \mid} -
\frac{GM_s}{\mid \bf{r} - \bf{D} \mid} +
{GM_s{\bf r} \cdot {\bf D}\over D ^3}
\label{tpot}
\end{equation}

\noindent where $G$ is the gravitational constant. The first dominant 
term is due to the primary with mass, $M_p,$ while the remaining perturbing terms are
due to the secondary with mass,  $M_s.$ The last indirect term accounts for the
acceleration of the origin of the coordinate system.

\noindent For secondary masses comparable to the primary mass it is convenient
to refer the description of an external disc to a coordinate system based
on the centre of mass of the binary. Then the gravitational potential
takes the form

\begin{equation}
\Psi = - \frac{GM_p}{\mid \bf{r} - \bf{D}_p \mid} -
\frac{GM_s}{\mid \bf{r} - \bf{D}_s \mid},
\label{tpotcom}
\end{equation}

\noindent where ${\bf{D}}_s=M_p{\bf{D}}/(M_p+M_s)$ and
${\bf{D}}_p=-M_s{\bf{D}}/(M_p+M_s).$ Note that in this case because the
origin is not accelerating there is no indirect term.

\noindent These potentials are used in the SPH calculations
with the minor modification that the potential due to a point
mass is softened. That is the standard Newtonian potential due to a point
mass $M$, namely
$-GM/r$, is replaced by $-GM/\sqrt{r^2+b^2}$, where $b$ is the softening
parameter, chosen to be small compared with the binary seperation.

\section{Viscous Evolution of the Disc Under Tidal Forcing}

\subsection{The Coplanar Case}

\noindent The evolution of binaries with coplanar circumbinary discs has
been studied previously using particle simulation methods
(see Lin and Papaloizou 1979, Artymowicz et al. 1991, Artymowicz
and Lubow 1994) and also finite diffference methods
(Lin and Papaloizou 1993).
 
\noindent The natural tendency in an isolated disc is for material to slowly
move radially inwards while angular momentum is transported outwards by the
action of viscosity.

\noindent When the central region of the disc is occupied by a
binary in a coplanar circular orbit, there is a tidal interaction between
the disc and binary, the latter not being a point mass.
Because the binary rotates faster than the circumbinary disc material,
angular momentum is transported from the orbit to the disc. If the viscosity
is not too large, inward radial migration of disc material is prevented and
a cavity or `gap' is maintained (Lin \& Papaloizou 1979).
 
\noindent The basic evolutionary effects on the disc may be found from
(\ref{diffe})
with the addition of an angular momentum source term to account for the angular
momentum injection resulting from the tidal effects of the secondary
(Lin \& Papaloizou 1986b, Pringle 1991).
Pringle (1991) found that the  disc surface
density profile became steadily shallower as the outer
boundary moved outwards to increasingly large radii on the viscous
timescale. For sufficiently strong injections of angular momentum, disc
material never penetrated interior to the inner radius  where the source was
located. Whether or not the inner cavity is maintained depends upon the rate
of angular momentum input due to tidal effects in comparison to its transport
due to viscous effects (see Goldreich \& Tremaine 1978, 1982,
Papaloizou \& Lin 1984, Lin \& Papaloizou 1993). For the case of small
$M_s/M_p,$ Lin \& Papaloizou (1986a, 1993) give the condition for maintainance
of the gap against viscous diffusion as
 
\begin{equation}
{M_s\over M_p} > {40\nu \over  D^2 \Omega }.
\label{cond}
\end {equation}
 
\noindent Those authors found this expression to be valid for
$M_s/M_p < 10^{-2}.$ It was not found to be applicable
to binary systems with a large mass ratio, where it would over--estimate
the strength of tidal torques in comparison to viscous diffusion.
 
\noindent In addition, for a gap to form in a disc with Mach number
${\cal M}$, we expect, for a low mass ratio, that the disc thickness should
not be more than the Roche lobe size associated with the secondary. This gives
the thermal condition
 
\begin{equation}
\left(\frac{M_s}{M_p}\right)^{1/3} >  \frac{1}{\cal M}.
\label{cond2}
\end{equation}

\subsection{The Non-Coplanar Case}

\noindent From the study presented in LNPT, as well
as very general considerations, we expect tidal effects to work to maintain an
inner cavity as in the coplanar case but at a somewhat reduced level when the
orbital and disc planes are inclined. However, we also expect the
phenomena of warping and twisting of the disc as well as disc precession to
occur in this case.
 
\noindent An initially large inner boundary radius of a circumbinary disc
shrinks because of viscous evolution until tidal effects become important
enough to halt it.
 
\noindent During this process the disc is also expected to develop a twisted
and warped structure. The timescale for setting this up is expected to be on
the order of the time the disc takes to communicate with itself over a scale
length comparable to the inner boundary radius either through sonic or viscous
effects. This is not greater than the viscous diffusion timescale on which the
disc contracts inwards. In addition, as a result of the net torque acting on
it because of the inclined binary orbit, precession of the disc is expected
to be induced.
 
\noindent In order to carry out an independent investigation
of the warping and precession of a circumbinary disc, we
consider the linear response of a disc to the component of the time averaged
perturbing potential which is mainly responsible for producing these effects.
We remark that because the SPH circumbinary discs are inevitably very
viscous the inner edge moves close to the binary orbit where it becomes
in general strongly perturbed. The quantitative applicability of a linear
response calculation is accordingly restricted to low mass ratios and small
orbital inclinations. However, we find that qualitative trends can be
reproduced in other cases.

\subsection{The Linear Secular Response due to  Binary with an Inclined Orbit} 

\noindent For the problem of an external disc we can consider an expansion
of (\ref{tpotcom}) in powers of $D/r$, taking the part with odd symmetry
with respect to reflection in the $z$-plane as the important component
contributing to a secular non-coplanar response (Papaloizou \& Terquem 1995).
So keeping terms up to second order
in $D/r$ and assuming a thin disc, ie. $z/r \ll 1$, we find the
zero--frequency component of the tidal potential for azimuthal mode number
$m=1$ to be the real part of $\Psi'\exp(i\varphi),$ where
 
\begin{equation}
\Psi^{\prime} = i\frac{3}{4}\frac{\Omega^{2}}{r}
\frac{M_sM_p}{(M_p+M_s)^2}D^{2}z
\sin (2 \delta).
\label{pexp}
\end{equation}
 
\noindent We consider the linear response of a thin disc to the perturbing
potential $\Psi'.$
 
\noindent Papaloizou and Lin (1995a) derived the equation
needed to describe the secular  warped structure taken on by a thin fluid
disc, which may have a {\it small} kinematic viscosity, when perturbed by a
non--coplanar binary companion in the form:
 
\begin{equation}
{d\over dr}
\left\{ {\mu\Omega^2\over
\left[ \Omega^2 {(1-i\alpha)}^2-\kappa^2 \right] } {dg\over dr} \right\} =
{i{\cal I} \over r},
\label{zerof}
\end{equation}
 
\noindent where
 
\begin{equation} {{\cal I}}  =
\int_{- \infty}^{+ \infty}\frac{\rho
z}{c^2_s}(\Psi'+2i\omega_p rz\Omega \sin \delta )dz .
\label{force}
\end{equation}
 
\begin{displaymath}
\mu = \int\limits_{+\infty}^{-\infty} \rho z^2 dz ,
\end{displaymath}
\begin{displaymath}
\kappa^2 = {2\Omega \over r}{d\over dr}(r^2\Omega) ,
\end{displaymath}
 
\noindent and we have allowed for a Shakura--Sunyaev $\alpha$
viscosity to act on the vertical dependence of horizontal motions. Note that
the effect of viscosity on the radial dependence
of the vertical displacement has been neglected. This might be important
in cases showing a strong non--linear response. It is likely
that this leads to the weaker response we find in our model discs than is
indicated by the linear theory described here. The complex function $g$ gives
the ratio of the vertical to rotational velocity $g\exp(i\varphi)$. In terms of
the vertical displacement $\zeta(r) \exp(i\varphi)$, $g=i \zeta /r$.

\noindent We suppose our coordinate system, defined as above
with respect to the disc at some initial epoch,
and hence the disc rotation axis, precesses about
the orbital rotation axis with angular velocity $\omega_p.$ This is
necessary (at least in the case of a finite disc)
because if the unperturbed disc density vanishes at
an inner boundary with $r=R_i$ and an outer boundary with $r=R_o,$ there is an
integrability condition associated with (\ref{zerof}) which is obtained by
integrating (\ref{zerof}) over the disc. This is
 
\begin{equation}
\int_{R_i}^{R_o}\int_{- \infty}^{+ \infty}{{\cal I}\over
r}dzdr=
\int_{R_i}^{R_o}\int_{- \infty}^{+ \infty}\frac{\rho z}{c^2_s}
{{(\Psi'+2i\omega_p rz\Omega \sin \delta )}\over
r}dzdr=0.
\label{INTG}
\end{equation}
 
\noindent The precesion frequency $\omega_p$ is chosen so that
the integrability condition (\ref{INTG}) is satisfied. This gives,
assuming a linear dependence for $\Psi',$
 
\begin{equation}
\omega_p=
-{\int_{R_i}^{R_o}{\Sigma\over r\Omega^2}\left({\partial
\Psi'\over \partial z}\right)_{z=0} dr\left/{\int_{R_i}^{R_o}
{{2i r\Sigma \sin(\delta)}\over\Omega
r}dr}\right. }
\equiv - \frac{3}{4}{GM_{s}M_{p} D^{2}\over (M_s+M_p)}\cos \delta
{\int {\Sigma r^{-2}} dr \over \int \Omega \Sigma r^3 dr} .
\label{WNTG}
\end{equation}
 
\noindent Here we have used
\begin{displaymath}
\int\limits_{+\infty}^{-\infty} \frac{\rho z^2}{c^2_s} dz =
\frac{\Sigma}{\Omega^2},
\end{displaymath}
an identity which follows by multiplying (\ref{vert}) by $\rho z/c^2_s$ and
integrating through the disc.
 
\noindent For a Keplerian disc, (\ref{WNTG}) gives the presession
frequency as the total torque acting on the disc divided by the total
disc angular momentum. For a finite disc this is always non zero. However,
for an infinite disc with infinite angular momentum content the precession
frequency is clearly zero. In such cases it is possible that the inner
regions adjust into a warped structure while exhibiting no precession,
even though there is a net torque on the disc (see below).
\noindent After having determined $\omega_p$ by the above procedures,
equation (\ref{zerof}) may be integrated to give
 
\begin{equation} g
=i\int^r_{R_i}{ [\Omega^2(1-i\alpha)^2-\kappa^2]\over
\mu\Omega^2}\left(\int_{R_i}^r{{\cal I}\over r}dr\right)dr
\label{zero1f}
\end{equation}
 
\noindent Here, we have assumed that $g=0$ for $r=R_i$ but note that an
arbitrary constant may be added to $g.$ This corresponds to a small
arbitrary rigid tilt (Papaloizou \& Lin 1995a) which we assume to be taken
care of by the choice of coordinate system. In order for the disc to
approximately precess as a rigid body $g$ calculated from (\ref{zero1f})
must be small.

\noindent As stated above, when the disc has infinite extent and angular
momentum content, $\omega_p$ is identically zero, although there is a net
torque on the disc. None the less (\ref{zero1f}) may give a
convergent expression for $g.$ Physically this corresponds to the situation
when the disc is distorted in the inner regions without undergoing any
global precession.

\subsection{The response of a disc of infinite extent}

We illustrate the possibility of solutions of this type
by considering polytropic models of the kind corresponding to our
simulations ($n=1.5$). We also make the replacement
$\Omega^2 =\kappa^2,$ and neglect $\alpha^2$ in comparison to $\alpha.$
This procedure is expected to be valid in the bulk of the disc
if $\alpha \ge (H/r)^2$ and also if $\alpha\Omega$ exceeds the precession
period of a particle orbit locally
(see for example Papaloizou \& Pringle 1983, Papaloizou \& Lin 1995a). These
conditions are  expected to hold for our simulations for
the smallest perturbing mass presented
below. The response equation behaves then like a diffusion equation. In this
case (\ref{zero1f}) gives for $\omega_p =0,$ and $\alpha$ assumed to be
constant

\begin{equation} g
=\frac{3i(2n+3)\alpha}{2}\sin
(2\delta)\frac{M_sM_p}{(M_p+M_s)^2}D^{2}\int^r_{R_i}{1 \over \Sigma
H^2}\left(\int_{R_i}^r   \frac{\Sigma}{r^2} dr\right)dr.
\label{zero2f}
\end{equation}
 
\noindent Equation (\ref{zero2f}) gives a convergent
expression for $g$ as $r\rightarrow \infty,$ provided
$1/( \Sigma H^2) $ vanishes sufficiently rapidly. For illustrative purposes,
we consider the case when $H/r=constant$
and $\Sigma \propto r^{-1/2}$ for $r>R_i,$ otherwise $\Sigma =0.$
But note that taking other power laws
for $\Sigma$ yields very similar results as long as the integrals are
convergent. In considering such models, we neglect the structure of the inner
disc edge, but as there is no divergence of the integrals there, this should
make little difference. For this simple model (\ref{zero2f}) gives
 
\begin{equation}
g=12i\alpha\sin (2\delta)\frac{M_sM_p}{(M_p+M_s)^2}
{\left(\frac{D}{R_i}\right)}^{2}
{\left({r \over  H}\right)}^{2}
\left[ {3\over 4} - \left({R_i\over r}\right)^{1/2} +
\left({R_i\over 2 r}\right)^{2} \right] .
\label{gee}
\end{equation}
 
\noindent The dependence on azimuth is such that the relative displacement,
$\zeta / r$, is maximised on the $x$ axis,
being an increasing function of $x$ for both $x>0$ and $x <0$. This behaviour
is similarly found to occur for the zero--frequency
response of a circumprimary disc (Papaloizou et al. 1995, LNPT).

\noindent We define the total range of $\zeta / r$ as
 
\begin{equation}
\Delta (\zeta / r)
=9\alpha\sin (2\delta)\frac{M_sM_p}{(M_p+M_s)^2}
{\left( {D \over R_i} \right)}^{2}
{\left({r \over  H}\right)}^{2}.
\label{geer}
\end{equation}
 
\noindent The condition for $\Delta (\zeta / r)$ to be small, and hence for the
linear analysis to have some validity, is, for $\alpha \sim
H/r,$ roughly that the sound crossing time
across a scale $\sim R_o$ be less than the inverse precession
frequency. For a very viscous disc, the criterion should be
that the viscous diffusion time be less than the inverse precession frequency
(Papaloizou \& Pringle 1983). But note that because the simulations are for
finite discs, they exhibit a non zero (albeit small) pseudo rigid body
precession rate. This makes essentially no difference to the above discussion
if the model is a cut off version of an infinite one with convergent
$\Delta (\zeta / r)$. However, it is in general possible for some models that
$\Delta (\zeta / r)$ is divergent. In this case the disc  may split into
disconnected pieces each of which has good enough internal communication
to maintain approximate rigid body precession. This kind of behaviour was
found to occur in very thin circumprimary disc models in LNPT (see also model
13, below).

\section{Numerical Simulations}

We have performed simulations for binary mass ratios, $q = M_s/M_p$ equal to
$0.01, 0.1,$ and $1.$ The range of Mach numbers considered is $10-30,$
and we take the inclination between binary orbit and initial disc plane
to be in the range $0 - 45$ degrees.

\noindent For $q=0.01$ we adopted a coordinate
system with the origin located at the primary (Case I). For the other
mass ratios, we adopted a coordinate system with the origin located at the
centre of mass of the binary (Case II). In each of these cases the $(x,y)$
plane of the fixed Cartesian coordinate system used to describe the disc
was taken to coincide with the initial disc mid--plane.

\subsection{Initial Conditions}

A disc containing 27000 identical particles was initially set in orbital
motion about the origin of the coordinate system. Each particle was then
 put into a
circular orbit, obtained assuming a unit mass,
acting as a softened  point mass, centred at the origin
(see above). The rotation law adopted for the disc was thus of the
modified--Keplerian form:

\begin{equation}
\Omega = {1 \over \left( r^2 + b^2\right)^{3/4}} .
\label{omeg}
\end{equation}

\noindent where $b$ is the softening length. 

\noindent We adopt units of mass such that for Case I, the primary mass
$M_p=1,$ and for Case II the total binary mass $M_s+M_p =1.$ 
In  both cases the gravitational constant $G=1,$ and the unit of length was
chosen such that the initial outer disc boundary was at $r=R_o =5.$
The softening length for the primary was taken to $b=0.2$ in these units. We
then adopt the natural time unit, being the inverse Keplerian frequency for unit
central mass at $r=1$. When $q=1$ the softening length for the secondary is
taken to have the same value as for the primary, when $q \ne 1$ we use
$b=0.001$ for the secondary softening.

\noindent The initial inner boundary of the disc was taken to be at
$r = R_i =1.5 $ in all cases with $q \ne 1.$  For $q=1,$ we took
$ R_i =2, $ the larger value being necessary because of the strong tidal
effects that occur when $q=1.$

\noindent  The particle positions were initialised by dividing the disc into
100 annulli of equal width and placing in each one an equal number of identical
particles  so as to obtain on average $\Sigma \propto 1/r.$ 
Initially 30 particles were placed at random in the mid--plane of
each annulus giving a disc plane containing a total of $3000$ particles.
 Eight other identical planes were then placed symmetrically about 
the mid-plane with equal vertical separation, such that the total initial disc
semi-thickness was $\bar H$. The  constant value of $\bar H$ was chosen such
that $R_o/{\bar H} \equiv {\bar {\cal M}},$ the
required Mach number (we shall subsequently use ${\bar {\cal M}}$ when we
refer to the single Mach number associated with a particular model). The
polytropic constant $K$ was adjusted so that the Mach number calculated in
the mid-plane at $r=R_o$ from the equation of state was  equal to the
required value. 

\noindent We remark that after relaxing to vertical hydrostatic equilibrium,
noting that the mid-plane sound speed is almost unchanged,
the total vertical semi-thickness
$H=\sqrt{3} \bar H$ at the outer boundary (see equation~(\ref{vertrho})). 
The mid-plane Mach
number is then found to be approximately independent of radius so that we can
effectively characterize each initial disc model by just one value.
We note that (\ref{vertsig}) implies that in the mid-plane ${\cal M} \propto
r^{1/8},$ being a very weak radial dependence.

\noindent For all of the calculations described here the smoothing length
turned out to be  less than the disc semi-thickness,
this indicates that the latter was
determined by genuine pressure effects and not kernel support, which is
necessary in order to model the hydrodynamics in three dimensions.

\noindent For $q=0.01$ and $q=1,$ the disc models were allowed to relax 
for approximately $2\Omega(R_{o})^{-1},$ evaluated at the initial
outer boundary, under the gravity of a central unit mass. 
During the initial relaxation period
the outside edge of the disc expanded by about 20 percent. This effect was
mostly due to a pressure-driven expansion into the surrounding vacuum. After
relaxation, the Mach number was found to be approximately constant through the
disc. Subsequently the full gravitational potential corresponding to a
binary in a circular orbit with separation $D=1$ was introduced. However, we
note that the relaxation time for vertical hydrostatic equilibrium
is short compared to the disc evolution timescales of interest.
Accordingly, after long times the results were found to be robust to
variation in the details of the initial conditions. For the results presented
with $q=0.1$, the full gravitational potential was introduced immediately
without prior relaxation. The orbital separation was taken to be $D=0.7$ for
these models. With this choice of $D$ for $q=0.1$ our different mass ratio
cases divide into three groups according to the relative strength of the
linear secular tidal potential.

\noindent The potential expansion we carry out allows us to make a
parameterisation for the `tidal strength'
$\varsigma \equiv \frac{M_sM_p}{(M_p+M_s)^2}D^{2}$. This parameter gives
a crude relative measure of the effectiveness of the tidal torques in
truncating a circumbinary disc, for fixed orbital inclinations. We also
consider the relative strength of the `warping tide',
$\varsigma \sin(2\delta)$. With this
choice of parameter the cases for mass ratios $q=0.01, 0.1$ and $1.0$
correspond to $\varsigma = 0.01, 0.05$ and $0.25$ respectively, so that an
increment in the $q$--value of the model modifies the tidal strength by a
constant multiplicative factor. We test our models for various values of
${\bar {\cal M}}$, $\varsigma$ and $\delta$ but for clarity we refer to $q$
in our discussion of the results, noting that the relative strengths of the
tidal torques at work for the respective models is related in the way
described above.

\subsection{Measuring Warping and Precession}

\noindent In order to measure the degree of warping and amount of precession
manifest in our calculations we introduce an angle $\iota$, being the
angle between the angular momentum vector of the disc material contained within
a specified cylindrical annulus,
and the total angular momentum vector of the disc. We define $\iota$ through
 
\begin{displaymath}
\cos \iota = { {\bf J}_A\cdot  {\bf J}_D \over | {\bf J}_A||{\bf J}_{D}|} .
\end{displaymath}
 
\noindent The disc angular momentum is ${\bf J}_D$, calculated as the
sum over all disc particle angular momenta.
The angular momentum
within an annulus is ${\bf J}_A$, calculated as the sum of all disc particle
angular momenta within the annulus. If $\iota$ is small its total range
is equivalent to the total range of $\zeta / r$, defined above.
  
\noindent The angle $\Pi$ measures the amount of precession of the disc
angular momentum vector, ${{\bf J}_D}$, about the binary orbital angular
momentum vector, ${{\bf J}_O}$. We define $\Pi$ through
 
\begin{displaymath}
\cos \Pi = { ({\bf J}_O \mbox{\boldmath $\times$}
{\bf J}_D)\cdot {\bf u} \over | {\bf
J}_O{\mbox{\boldmath $\times$} \bf J}_D| |{\bf u}|} .
\end{displaymath}
 
\noindent The reference vector ${\bf u}$ may be taken to be any
arbitrary vector in the orbital plane. For convenience we choose ${\bf u}$
such that $\Pi$ takes the initial value of
$\pi / 2$ radians in all models. For retrograde
precession of ${\bf J}_D$ about ${\bf J}_O$ the angle $\Pi$ should
decrease linearly with time as is found in practice.

\section{Numerical Results}

Before describing the results of our simulations in detail, we give a brief
summary of our findings. In all cases the initial location of the disc inner
boundary was too great for angular momentum input from tidal torques to
balance inward viscous diffusion. All the coplanar models except those with
$q=0.01$ and ${\bar {\cal M}}=10$ reached a quasi-equilibrium in which a
central cavity containing the binary was essentially maintained, with further
disc contraction through viscosity being halted because of the angular momentum
input by tidal torques. The gap--clearing efficiency of tidal torques
appeared to become reduced as the inclination of the disc plane to that of
the orbit increased. This was apparent due to the reduced size of the cavity
at higher inclinations of the companion's orbit compared with the greater
scale of the cleared cavity present for lower inclinations. When $q=0.01$
and ${\bar {\cal M}}=10,$ the viscous condition for tidal
truncation (\ref{cond}) gives $\alpha / {\cal M}^2 < 2.5\times 10^{-4},$
which is not satisfied near the inner disc boundary. Accordingly a breakdown
of tidal truncation is observed in these cases. 

\noindent The tidally truncated inclined discs also exhibit warping.
For mass ratios $q=0.01$ and $q=0.1,$ we found that
the magnitude of the elevation of the disc mid--plane above its initial level
behaved qualitatively as predicted in Section~(4) above. In addition, except
when $q=1,$ very slow quasi-rigid body precession was observed. These cases
also implied a very slow evolution (possibly on the global viscous
timescale) of the relative mean disc plane inclination to the orbital plane
(cf. Papaloizou \& Terquem 1995).

\noindent When $q=1$ the tidally truncated inclined discs were 
severely warped in comparison to the other models. However, a 
quasi--rigid body precession was still observed for small inclinations
such that the disc maintained itself as a unit. For high inclinations
the disc distortion became so severe that there were indications that
it could evolve into disconnected annuli. Parameters for all the models we
consider are presented in Table~(\ref{table1}).

\subsection{Coplanar Models}

\noindent We begin with a discussion of our results obtained when
the disc and orbital planes were coincident. These cases, which may be
related to previous work, provide reference points to which the non coplanar
cases may be compared (see Lin \& Papaloizou 1993, and references therein,
and also Artymowicz \& Lubow 1994).

\subsubsection{Models with q=0.01}

\noindent {\it Model 1}

\noindent This model had ${\bar {\cal M}}=20$. Figure~(\ref{fig1}) shows a
particle projection plot onto the $(x,y)$ plane
after relaxation was complete, but without a binary companion ($M_s =0$).
This is to be compared to Figure~(\ref{fig2}) which shows a similar projection
plot of the disc after the companion had been included for an additional time
of approximately 150 units. This shows a cavity to have been cleared out to as
far as a radius $\simeq 1.4$. A faint density wave with $m=1$ is clearly seen,
as well as a `wake--like' feature trailing the companion, causing the shape
of the inner boundary to deviate from axisymmetry.

\noindent At this stage the inner boundary region of the disc appears to
have reached a quasi-steady state with tidal interaction providing a
continuous supply of angular momentum to balance viscous effects, which
otherwise would cause material to flow inwards. This is consistent with the
viscous condition for tidal truncation (\ref{cond}) which, being marginally
satisfied at the inner disc boundary, implies that tidal truncation should
just be possible.

\noindent However, a few particles were found to move inside the cavity
as a material stream reached toward the companion from the inner boundary.
This leakage is possibly due to poor statistics combined with the effects of
large smoothing lengths near the inner boundary. Due to these effects, small
numbers of particles could be propelled into unstable coorbital trajectories.
The numbers of particles involved is too small to make any inference about
consequent local fluid properties. We comment that, for this to be possible,
a more thorough understanding of the behaviour of SPH particles in low
density regions and near edges than appears to be currently available is
required.

\noindent As the `wake--stream' of particles extends from the inner boundary
to the companion (see Fig.~(\ref{fig3})), its inner edge becomes increasingly
poorly defined. It is not clear what role numerical effects play in its
existence. Although, this feature of our numerical results can be
understood in terms of a simple impulse model for tidal interaction (Lin \&
Papaloizou 1985). It is plausible that, if the companion mass were to be
reduced, the wake--stream may ultimately tend to the streamline configuration
seen in the vacinity of an accreting protoplanet fully embedded in a thin
disc (Korycansky \& Papaloizou 1996).

\medskip
\noindent {\it Model 6}

\noindent We describe this model here as it had the same
parameters as model 1 except that the Mach number was halved by comparison. 
This had the consequence that the effective shear viscosity was approximately
two and a half times larger (see equation (\ref{visc})). In this case the
condition (\ref {cond}) implies a failure of tidal truncation near the inner
disc boundary. In model 6 the inner disc is found to approach the companion's
orbit more closely than in model 1. This is consistent with the disc material
seeking a stronger tidal torque to balance the increased viscous transport.

\noindent Indeed the inner boundary for this model contracted to a mean
radius of approximately $1.2$, which is about a companion Roche lobe radius
away from the coorbital radius. Thus the inner boundary had closed $0.2$ radius
units, which is more than one would expect from thermal effects
alone. As expected the large viscosity in these low mass ratio models
determines the size of the gap.

\noindent When the disc inner edge had approached so
close to the coorbital radius that non--linear tidal disruption of the inner
boundary occurred, many particles were found to move interior to the binary
orbit in an apparently continuous transfer of material. The mean rate of
flow of particles into the inner region
was found to be comparable to the rate of delivery of material expected
through unimpeded viscous transport. However, the number was still
too small for detailed fluid properties to be represented beyond saying that
the inflow did occur.

\noindent Figure~(\ref{fig4}) shows a projection plot onto the initial disc
mid--plane for model 6 after a time of about 125 units after the introduction
of the secondary. Any particles contained within a radius of unity of the
computational origin were extracted at a time of about 75 units.
Therefore, in this plot, all the particles interior to the coorbital radius
($\sim 100$ of them) found their way there over a period of about $50$ time
units. We note that the character of the gap breakdown in this marginal case
is to maintain the cavity, allowing leakage through the vacinity of the
companion's position without it `accreting' any material.

\noindent The gap opening criteria (\ref{cond}) \& (\ref{cond2}) are found to
be consistent with our results, confirming our calibration of the disc
viscosity. It is a consequence of the scaling for the shear
viscosity (\ref{visc}) that the disc thickness and viscosity are not
independently variable as model parameters.

\subsubsection{Model with q=1}

\noindent By contrast with the marginal case of model 6 above, the unit mass
ratio model 11 with ${\bar {\cal M}}=20$ displayed strong tidal torques
with no indication of any particle leakage into the interior or the development
of wake-streams. Results were found to match expectation
(Lin \& Papaloizou 1979, Artymowicz \& Lubow 1994).
A cavity was cleared out to a mean distance of about $1.9$ units, measured
from the binary centre of mass.

\subsection{Models with non--zero inclination}

\noindent As the inclinations were increased in all our models,
the incidence of wake-streams and gap leakage increased also. This is
clearly due to the weakened tidal torques at higher orbital inclinations
allowing the closer approach of the inner boundary to the binary components.
This highlights that the formation of wake-streams depends on the relative
strengths of the tidal and viscous torques, so that such features are expected
to occur only when viscosity predominates. Furthermore,
the gap edge in low inclination, low Mach number (i.e.
large viscosity) models showed similar behaviour to that for higher Mach
number (i.e. smaller viscosity), higher inclination cases. 

\noindent We now proceed to discuss the response of our
models with inclination different from zero.

\subsubsection{Models with q=0.01}

Taking the final state of model 6 as the set of initial conditions (see
Figure~(\ref{fig2})), we introduced an inclination of the secondary's orbit
to the mid-plane of the disc, namely the $(x,y)$ plane. The inclination,
$\delta$, was increased in steps of $10$ degrees, each change being initiated
after an additional relaxation period of approximately 150 time units. For
each value of $\delta,$ the disc was evolved for several hundred time units
enabling the disc edge region and warped structure to achieve a quasi--steady
state.

\noindent In general the degree of warping was mild and much less than would
be expected from a collisionless particle simulation (see below). Rough
agreement was obtained with the simple linear theory described in Section 4,
at least when the degree of warping was small, as in model 2. In addition we
observed in such cases that the disc behaved in a manner consistent with
approximate rigid body precession. However, the precession frequencies were
small so that total angles of precession were small. Thus this effect was not
dynamically very significant and as we have indicated above would probably
disappear if the disc were to have arbitrarily large extent.

\noindent As discussed above, the introduction of a non--zero
inclination results in a reduction in the effectiveness of the tidal
torques in maintaining a cleared gap. This results in the disc inner edge
moving inwards in comparison to the coplanar case for all models.
Consequently tidal truncation could break down at a finite inclination.

\medskip
\noindent {\it Models 2--5}

\noindent Figure~(\ref{extra}) shows a projection plot of the model 4
data at a time of $265$ units after the introdution of the companion on
an inclined orbit at $30$ degrees to the initial disc mid--plane. The
gap generally remains cleared, suffering only a small contraction of the disc
inner edge. Also a wake--stream seems to have become a continuous feature in
this case, as compared with its episodic growth and decay in the coplanar
case of model 1. The rate of inflow of particles inside of the projected
orbital path is still very slight though, being much less than that
found in the coplanar model 6. We find that the gap is maintained
for a finite orbital inclination of the companion for all of the models 2--5,
with indications that the gap would breakdown if the inclination were much
higher.

\noindent The precession angle, $\Pi$, evolution for all of these models
is plotted in Figure~(\ref{fig5}) and the precessional timescales
(or inverse precession frequencies) that we infer,
$\langle{\omega_{p}^{-1}\rangle}$, are given in
the second column of Table~(\ref{table2}), for each value of $\delta$.
Typical values are in broad agreement with
equation~(\ref{WNTG}); e.g. assuming $R_{i}/R_{o} = 1.2/6$,
$\Sigma \propto 1/r$, Keplerian rotation and $\delta= 20$ degrees,
we find $\omega_{p}^{-1} \simeq 3000$. None of these models showed precession
through more than about 10 degrees. As we shall see, when the strength of the
tidal force increases with the larger mass ratio cases, the disc precession
can be of much greater significance. 

\noindent Figure~(\ref{fig6}) plots $\iota$ as a
function of radius for model 5 with $\delta=40$ degrees, the inclination
profile appears to reach a state where it changes very slowly with time. 
We remark that some residual evolution would be expected on a viscous
timescale as the disc density decreases in the vicinity of the binary orbit.
The data implies a range of $\zeta / r \simeq 0.14$ in this case, for model 2
with $\delta=10$
we infer $\Delta (\zeta / r) \simeq 0.06$. This is in reasonable agreement with
what we obtain using (\ref{geer}); for the typical values $M_s/M_p =0.01,$
$\alpha =0.04,$ ${\cal M}=20,$ $R_i/D =1.5,$ we find
$\Delta (\zeta / r)  \simeq0.21 \sin (2\delta)$. Then for $\delta = 10$ and
$40$ degrees, $\Delta (\zeta / r) \simeq 0.07$ and $0.21$ respectively (note
that we do not make any notational distinction between inferred and predicted
values). Model 2 gives the better agreement, as might be expected considering
the weaker warping tide in this case. In view of the uncertainties involved in
the details of the edge region and the $\Sigma$ profile to be used we
consider the general qualitative agreement that we find for these models to
be satisfactory.

\medskip
\noindent {\it Models 7--10}

\noindent Performing similar calculations for a lower Mach number such as in
models 7--10, results in a lower set of values for
$\Delta (\zeta / r)$ in all cases (as we expect from
considering a disc with a larger sound speed). For $\delta = 10$ and $40$
degrees we expect $\Delta (\zeta / r) \simeq 0.013$ and $0.039$ respectively.
We infer from the data the values $0.011$ and $0.018$ which shows once again
that the best agreement is for the weakest warping tide (i.e. smallest value
of $\delta$) and is poorest when the warping tide is strongest. The agreement
is generally poorer in this case though, probably due to non--linearity and
excessive gap leakage occurring near the inner edge. The precession
frequencies we infer, however, are hardly altered by comparison with those for
models 2--5. This shows that the global secular response is determined by
communication over the whole disc and the inner warp is set up locally. 

\noindent Agreement with our analytical expressions matches well in these
cases as it did for models 2--5. The only significant difference is the
extent of the gap leakage, which is larger at low inclinations than
in the higher Mach number models, becoming increasingly important at larger
inclinations of the companion's orbit.

\subsubsection{Comparison with a disc of non--interacting particles}
 
\noindent We now highlight the striking difference in the response of a fluid
disc with hydrodynamic forces and a disc of non--interacting particles.
The response of such a ballistic model to the tide of a companion on
an inclined orbit is to warp the disc by the propagation of a kinematic
bending wave from the innermost parts to the outermost parts of the disc.
This reaches a radius where the  precessional timescale for non--interacting
orbital planes is roughly equal to the time after initiation (Mouillet et
al. 1996). For small inclinations the scale of the warp is typically on the
order of the maximum vertical displacement of the companion and the disc does
not precess as a single entity.
 
\noindent Figure~(\ref{fig7}) shows a comparison between a ballistic
model and model 5 at similar times. The ballistic model was setup by
randomising particle positions within a fixed opening angle to obtain the
appropriate aspect ratio for direct comparison with the hydrodynamic model.
The SPH disc in model 5 shows a damping of the large warp present in the
ballistic disc at short times. In this case the warp amplitude is affected by
communication with the outer parts through pressure waves on a short timescale
and viscous effects on a much longer timescale. The vertical scale of the inner
warp is clearly much smaller than in the ballistic case. Further, the scale
of the warp is found to approach a quasi--steady configuration with this
small magnitude, as can be seen in Figure~(\ref{fig6}). 

\noindent At long times the ballistic disc became generally thickened on the
scale of the original inner warp. The ballistic disc was not found to show
global rigid--body precession on any timescale.

\subsubsection{Models with q=0.1}

\noindent Models 14 \& 15 were initialised with ${\bar {\cal M}}=20$,
and model 16 with ${\bar {\cal M}}=30.$ In each of these cases the
binary separation was taken to be $D=0.7$. In Model 14 the orbit was given an
initial inclination to the disc of $\delta = 10$ degrees, and in models 15
and 16 this was taken to be
$\delta = 45$ degrees. For these models the companion was introduced directly
without prior disc relaxation. The sudden introduction of the binary resulted
in transient axisymmetric waves which propagated and damped on the
sound--crossing timescale. The disc inner boundary spread
inwards until it became truncated at $r \simeq 1$ with an inner cavity being
maintained for the duration of the runs.

\noindent In spite of non--linearity, $\zeta / r$ has a similar form to
that in the lower mass ratio cases, with the maximum values occuring along
the $x$ axis. We show this is in Figure~(\ref{fig8}) by comparing a
`sectional--plot' (in which we only consider particles such that
$-0.5 < x,y < 0.5$) of the disc as seen projected onto the $(x,z)$ and
$(y,z)$ planes for model 14 after a time of about 310 units.

\noindent The longest run for the case of $q=0.1$ was for model 16, with a
run time of 410 units. Note that because the binary separation is $D=0.7,$
this corresponds to 700
inverse binary frequency time units. After this time only slow evolution of
the warp was observed as in the $q=0.01$ case.
In order to illustrate the behaviour of the warp as a function of Mach number,
we give $\iota$ versus radius plots for models 15 and 16 at the same time
of 310 units in Figure~(\ref{fig9}). This demonstrates that the warp amplitude
is less for the lower Mach number case. 

\noindent For  cases with $q=0.1,$  we infer an extremely long quasi--rigid
body precessional timescale $\sim 10^{4}$ time units. This is probably
because relatively little of the disc matter finds its way to small radii
as a result of the large tidal angular momentum input, yielding a small net
precessional torque. In all of these cases, the larger part of the disc
essentially maintained its original plane over the period of the simulations.

\subsubsection{Models with q=1.0}

These unit mass ratio cases show the strongest warping perturbations and are
therefore instructive in understanding the way that a circumbinary disc
responds to warping perturbations. Figure~(\ref{fig10}) shows sectional--plots 
for the model 12 disc at times of about $250$ and $450$
units after the introduction of the companion on a path inclined at
$\delta=10$ degrees (after the coplanar relaxation of model 11). In the first
frame we see that the warping has its largest
affect in the inner regions of the disc, with an annulus of material lifting
out of the original plane. Consistent with our choice of surface density
profile and equation~(\ref{zero2f}), the annulus does not
seperate from the outer regions of the disc but, as shown in the second
frame, settles together with the outer parts to give a stable global warp.

\noindent Figure~(\ref{fig11}) shows the precession angle evolution for the
duration of this run. The disc exhibits uniform precession through an angle of
approximately $180$ degrees and we infer
$\langle \omega_{p}^{-1} \rangle \simeq 300$ time units, consistent 
to  within a factor of two with equation~(\ref{WNTG}), which yields
$\omega_{p}^{-1} \simeq 180$ (assuming $\Sigma \propto 1/r$,
$R_i / R_o = 1.9 / 6$ and Keplerian rotation). 

\noindent Taking the relaxed configuration of model 12, we incremented the
inclination to $\delta = 30$ degrees. This resulted in an extreme non--linear
response of the disc such that the disc became severely warped and a dense
ring of material gathered at the inner boundary. This redistribution of the
surface density enabled the ring to partially disconnect from the outer disc.
The disconnection results in two separate inclined discs which
try to precess separately (see also LNPT) but which have a strong interaction
region where material from the two discs grazes. It appears that non--linear
dissipation has occured in a shocked region where material from the almost
disconnected discs interact. This interaction attempts to bring the discs
back to synchronous precession. The simulations we have carried out indicate
that structures like this can be long lived. In a real system this effect might
produce an observational signature as a ring of emission.
Figure~(\ref{fig12}) displays the position data resulting from this simulation
in a particle projection onto the $(x,y)$ plane. Figure~(\ref{fig13}) takes a
projection at a viewing angle such
that the outer disc is seen almost edge--on. In this figure, an inner ring is
clearly seen at a small inclination to the outer disc.

\section{Discussion}

\subsection{Tidal Truncation}

\noindent As was found for circumprimary discs in LNPT, provided
the viscosity is small enough, circumbinary discs are truncated at the radius
where viscous and tidal torques balance. As far as it can be investigated
with our SPH method, gap clearing in a disc by an embedded companion on a
fixed circular orbit is effective in discs with vertical structure and
consistent with theoretical expectation, both in the coplanar and inclined
orbit cases (Lin \& Papaloizou 1993, LNPT). The effect of the tidal torques
that clear a gap in the disc becomes weaker as the disc plane and binary orbit
become more inclined, with the consequence that the disc inner boundary moves
inwards and that the gap may breakdown at a finite inclination.

\noindent However, it should be emphasized that for the SPH models considered
here, the viscosity is inevitably large corresponding to Reynolds' numbers
typically $< 10^4$. In addition the disc thickness and viscosity are not
independently variable as model parameters. The effective shear viscosity
increases with disc thickness so that the viscous torques increase in their
effectiveness, in addition to the improved sonic communication. Thick discs
with very small viscosity or weakly viscous discs truncated by low mass
companions cannot be studied by this method.

\noindent When the viscosity is too large for its effects to be balanced by
tidal torques, an approximate inner boundary of the disc may be maintained
while particles spill into the gap at a significant rate through material
streams. This could be potentially important with regard to planetary
formation and accretion in the early solar nebula, but note that the
$\alpha \sim 0.03$ viscosity parameter, for which we mostly observe such
effects, is larger than those inferred for protostellar discs from estimates
of disc lifetimes (Beckwith et al. 1990), or obtained so far from
magnetohydrodynamical simulations (e.g. Brandenburg et al. 1996). 
But note values of $\alpha \sim 0.03$ have been found for unstable
self--gravitating discs (see Papaloizou \& Lin 1995b, and references therein).

\subsection{The Affect of Non--Coplanarity on Circumbinary Discs}

\noindent Provided sonic or viscous communication over a scale length 
comparable to the inner boundary radius takes less than the local inverse
precession frequency, and the surface density is not a very steep power law of
radius, the inclined disc responds to the secular torques
by precessing approximately as a rigid body. The precession frequencies were
seen to be small in the $q=0.01$ and $q=0.1$ cases. It is likely, as
demonstrated in a linear response calculation, that were the disc to extend
to an arbitrarily large radius, the precession frequency would be zero.
In addition to precession the disc takes on a warped structure in the
vicinity of the disc inner boundary, which in general has the same
qualitative form as that given by our linear response calculation.
For the $q=1$ case and high inclination, the disc apparently
could not maintain itself against the strong differential precession,
there being evidence of separation into two separate sonically connected
annuli. In this state the disc gave the appearance of a highly warped state.

\noindent Finally we comment that the response of discs with pressure and
viscosity is qualitatively and significantly different
from that seen for non--interacting particle discs.

\section*{Acknowledgments}

This work was supported by PPARC grant GR/H/09454, JDL is
supported by a PPARC studentship and is grateful to Vasso Agapitou
for a critical reading of the manuscript.


\section*{References}

\noindent Artymowicz P., Clarke C.J., Lubow S.H., Pringle J.E., 1991, ApJ,
370, L35 \\

\noindent Artymowicz P., Lubow S.H., 1994, ApJ, 421, 651 \\

\noindent Artymowicz P., Lubow S.H., 1996, ApJL, {\it submitted} \\

\noindent Bally J., Devine D., 1994, ApJ, 428, L65 \\

\noindent Beckwith S.V.W., Sargent A.I., Chini R., G\"usten R., 1990, AJ, 99,
924 \\

\noindent Brandenburg A., Nordlund \AA., Stein R.F., Torkelsson U., 1996,
ApJ, 458, L45 \\

\noindent Clarke C.J., Pringle J.E., 1993, MNRAS, 261, 190 \\

\noindent Corporon P., Lagrange A.M., Beust H., 1996, A\&A, 310, 228 \\

\noindent Dutrey A., Guilloteau S., Simon M., 1994, A\&A, 286, 149 \\
 
\noindent Goldreich P., Tremaine S., 1978, Icarus, 34, 227 \\
 
\noindent Goldreich P., Tremaine S., 1982, ARA\&A, 20, 249 \\
 
\noindent Heller C., 1993, ApJ, 408, 337 \\
 
\noindent Jensen E.L.N., Mathieu R.D., Fuller G.A., 1996, ApJ, 458, 312 \\
 
\noindent Korycansky D.G., Papaloizou J.C.B., 1996, ApJS, 105, 181 \\

\noindent Larwood J.D., Nelson R.P., Papaloizou J.C.B., Terquem C., 1996,
MNRAS, 282, 597 \\

\noindent Lin D.N.C., Papaloizou J.C.B., 1979, MNRAS, 186, 799 \\

\noindent Lin D.N.C., Papaloizou J.C.B., 1985, in Black D.C., Matthews M.S.,
eds, Protostars and Planets II (Univ. Arizona Press, Tucson) p.981 \\

\noindent Lin D.N.C., Papaloizou J.C.B., 1986a, ApJ, 307, 395 \\

\noindent Lin D.N.C., Papaloizou J.C.B., 1986b, ApJ, 309, 846 \\

\noindent Lin D.N.C., Papaloizou J.C.B., 1993, in Levy E.H., Lunine J., eds,
Protostars and Planets III (Univ. Arizona Press, Tucson) p.749 \\

\noindent Lynden-Bell D., Pringle J.E., 1974, MNRAS, 168, 603 \\

\noindent Mathieu R.D., 1994, ARA\&A, 32, 465 \\

\noindent McCaughrean M.J., O'Dell R.C., 1996, AJ, 111, 1977 \\

\noindent Monaghan J.J., Gingold R.A., 1983, J. Comput. Phys., 52, 374 \\

\noindent Monaghan J.J., 1992, ARA\&A, 30, 543 \\

\noindent Mouillet D., Larwood J.D., Papaloizou J.C.B., Lagrange A-M., 1996,
{\it submitted} \\

\noindent O'Dell C.R., Wen Z., Hu X., 1993, ApJ, 410, 696 \\

\noindent Nelson R.P., Papaloizou J.C.B., 1993, MNRAS, 265, 905 \\

\noindent Nelson R.P., Papaloizou J.C.B., 1994, MNRAS, 270, 1 \\

\noindent Papaloizou J.C.B., Pringle J.E., 1983, MNRAS, 202, 1181 \\

\noindent Papaloizou J.C.B., Lin D.N.C., 1984, ApJ, 285, 818 \\

\noindent Papaloizou J.C.B., Korycansky D.G., Terquem C., 1995, Annals of
the New York Academy of Sciences, 773, 261 \\

\noindent Papaloizou J.C.B., Lin D.N.C., 1995a, ApJ, 438, 841 \\

\noindent Papaloizou J.C.B., Lin D.N.C., 1995b, ARA\&A, 33, 505 \\

\noindent Papaloizou J.C.B., Terquem C., 1995, MNRAS, 274, 987 \\

\noindent Pringle J.E., 1981, ARA\&A, 19, 137 \\

\noindent Pringle J.E., 1991, MNRAS, 248, 754 \\

\noindent Roddier C., Roddier F., Northcott M.J., Graves J.E., Jim K., 1996,
ApJ, 463, 326\\
 
\noindent Shakura N.I., Sunyaev R.A., 1973, A\&A, 24, 337 \\
 
\noindent Terquem C., Bertout C., 1993, A\&A, 274, 291 \\

\noindent Terquem C., Bertout C., 1996, MNRAS, 279, 415 \\



\newpage

\begin{table}
\caption{Model parameters.}
\begin{center}
\begin{tabular} {ccccccccccc} \hline \hline

Model & ${\bar {\cal M}}$ & $q$ & $D$ & $\delta$ & $\varsigma \sin (2\delta)$ 
\\ 

\hline

 1 & 20 & 0.01 & 1.0 &   0 & 0.000 & \\
 2 & 20 & 0.01 & 1.0 &  10 & 0.003 & \\
 3 & 20 & 0.01 & 1.0 &  20 & 0.006 & \\
 4 & 20 & 0.01 & 1.0 &  30 & 0.009 & \\
 5 & 20 & 0.01 & 1.0 &  40 & 0.010 & \\
 6 & 10 & 0.01 & 1.0 &   0 & 0.000 & \\
 7 & 10 & 0.01 & 1.0 &  10 & 0.003 & \\
 8 & 10 & 0.01 & 1.0 &  20 & 0.006 & \\
 9 & 10 & 0.01 & 1.0 &  30 & 0.009 & \\
10 & 10 & 0.01 & 1.0 &  40 & 0.010 & \\
11 & 20 & 1.0  & 1.0 &   0 & 0.000 & \\
12 & 20 & 1.0  & 1.0 &  10 & 0.086 & \\
13 & 20 & 1.0  & 1.0 &  30 & 0.217 & \\
14 & 20 & 0.1  & 0.7 &  10 & 0.017 &\\
15 & 20 & 0.1  & 0.7 &  45 & 0.049 & \\
16 & 30 & 0.1  & 0.7 &  45 & 0.049 & \\

\hline \hline
\end{tabular}
\end{center}
\label{table1}
\end{table}

\begin{table}
\caption{Inferred precessional timescales for models 2--5.}
\begin{center}
\begin{tabular} {ccccc} \hline \hline

$\delta$ & $\langle{\omega_{p}^{-1}\rangle}$ \\ 
\hline

 10 & 2300 & \\
 20 & 2700 & \\
 30 & 3400 & \\
 40 & 4400 & \\

\hline \hline
\end{tabular}
\end{center}
\label{table2}
\end{table}

\newpage

\begin{figure}
\caption{The relaxed disc without a companion. Particle positions are projected
onto the $(x,y)$ plane. The sense of positive rotation is
anti--clockwise.}
\label{fig1}
\end{figure} 

\begin{figure}
\caption{The inner regions of the disc, relaxed with a companion included for
$\simeq 150$ time units. The primary is
shown as an asterisk, the companion as a circle and its projected path as a
dashed line. For clarity, this plot shows only the inner regions of the disc.}
\label{fig2}
\end{figure} 

\begin{figure}
\caption{The inner disc with a 'wake--stream' extending from the disc inner
edge to the direction of the companion after a further binary orbital period.}
\label{fig3}
\end{figure} 

\begin{figure}
\caption{The thicker more viscous disc of model 6 showing gap leakage
at a time of approximately $125$ units after the introduction of the
companion. We note also that the particles inside the cavity have not
suffered large vertical displacements and are therefore representative of gap
leakage.}
\label{fig4}
\end{figure} 

\begin{figure}
\caption{The thin disc of model 4 with the companion at an inclination
of $30$ degrees. Data is shown at a time of approximately $265$ units after
incrementing the companion's inclination. The cavity was cleared of
particles at a time of $140$ units. We note also that the particles inside
the cavity have not suffered large vertical displacements and are therefore
representative of gap leakage.}
\label{extra}
\end{figure} 

\begin{figure}
\caption{The precession angle versus time for models 2--5, shown for a
representative subset of the SPH data.}
\label{fig5}
\end{figure} 

\begin{figure}
\caption{The time evolution of disc inclination versus radius for model 5.
Note that the relative inclination angle plotted is essentially equivalent
to $| \zeta / r |$. For clarity we only plot data at intervals of approximately
$200$ time units.}
\label{fig6}
\end{figure} 

\begin{figure}
\caption{We project particle positions onto the $(x,z)$ plane. The lower frame,
for the ballistic data, is to be compared with the SPH data in the upper frame;
each is taken at a time of approximately $300$ units.
Notice that the fluid disc has taken on a mild
global warp, the ballistic disc is only warped locally, at this time, and by
a much greater amount.}
\label{fig7}
\end{figure} 

\begin{figure}
\caption{Sectional projection plots for model 14 at a time of approximately
$300$ units show that the warp is largest along the $x$--axis, 90 degrees out
of phase with the maximum of the warping potential.}
\label{fig8}
\end{figure}

\begin{figure}
\caption{Disc inclination versus radius plots for models 15 and 16, each at a
time of approximately $310$ units.}
\label{fig9}
\end{figure} 

\begin{figure}
\caption{Sectional projection plots for model 12 at times of 250 (upper frame)
and 450 (lower frame) units are shown.}
\label{fig10}
\end{figure}

\begin{figure}
\caption{The precession angle evolution for model 12. We plot the
magnitude of negative values.}
\label{fig11}
\end{figure}

\begin{figure}
\caption{Projection onto the $(x,y)$ plane for model 13 at a time of 730
units.}
\label{fig12}
\end{figure}

\begin{figure}
\caption{Projection of model 13 data at a time of 730 units, taken
at a viewing angle so that the outer disc appears approximately edge--on.}
\label{fig13}
\end{figure}

\end{document}